# A Fault Analytic Method against HB$^+$

José Carrijo, Rafael Tonicelli, and Anderson C. A. Nascimento

*Abstract*—The search for lightweight authentication protocols suitable for low-cost RFID tags constitutes an active and challenging research area. In this context, a family of protocols based on the LPN problem has been proposed: the so-called HB-family. Despite the rich literature regarding the cryptanalysis of these protocols, there are no published results about the impact of fault analysis over them. The purpose of this paper is to fill this gap by presenting a fault analytic method against a prominent member of the HB-family: HB$^+$ protocol. We demonstrate that the fault analysis model can lead to a flexible and effective attack against HB-like protocols, posing a serious threat over them.

## I. Introduction

**HB-family.** Recently, radio frequency identification (RFID) systems have attracted substantial attention from the industry and the research communities. Since RFID tags are expected to replace traditional barcodes, they will become one of the most used devices in the near future. They present a wide collection of applications that include: supply chain management, warehouse inventory control, pet identification, secure passport systems, anticounterfeiting tags for pharmaceutical products, among others.

Despite the benefits that could been brought by the use of RFID tags, some challenging issues emerge. The tag-reader communication occurs by means of a wireless link, what increases the system's vulnerability against passive and active adversaries. Additionally, RFID tags are highly constrained devices and do not have the processing, storage, power and communication resources necessary to implement standard authentication protocols.

In order to provide lightweight authentication functionalities to RFID systems, Hopper and Blum [9] introduced the HB protocol, which was based on a well-known intractability assumption: the Learning Parity with Noise (LPN) problem. The LPN problem is NP complete and only involves the use of binary vectors and inner products, what makes it an appropriate choice for constrained devices. Later, Juels and Weis [11] proved that HB was insecure against active adversaries and proposed an improved version of the protocol: the HB$^+$ protocol. However, Gilbert *et al.* [8] demonstrated that HB$^+$ was vulnerable to a class of man-in-the-middle attacks known as GRS-MITM attacks. In pursuance of mitigating this vulnerability, subsequent variants have been proposed: HB$^{++}$ [4], HB* [5], and HB-MP [12]. Unfortunately, neither of them succeeded. Their tolerance to GRS-MITM attacks is equivalent to that of HB$^+$ and they possess additional complexity and/or reduced practicality [7]. In 2008, Gilbert *et al.* [6] presented their protocol version called HB$^\#$. This new variant solved many drawbacks of its predecessors. Among them: it is provably secure against GRS-MITM attacks and presents a reduced communication cost. Although HB$^\#$ represented an evolution in terms of efficiency and security, it was later shown that it is vulnerable to a more general man-in-the middle adversary [13].

As illustrated in the previous history line, the research community effort has been directed to mitigating active attacks based on a man-in-the-middle setting. Before proceeding further, it is relevant to discuss the practical feasibility of man-in-the-middle attacks against the HB-family. No implementation of such attacks on RFID systems has been reported in the literature yet as it apparently requires a sophisticated hardware device capable of capturing and modifying the tag-reader communications in real-time [14]. In this article, we offer an alternative approach: a simple and effective fault attack that can be applied to a wide collection of HB variants. In contrast to MITM attacks, our approach has already been demonstrated to be feasible in [10], where implementations of fault analytic techniques over RFID tags have been made with low-cost equipment.

**Fault Analysis.** Traditional cryptography often assumes that secrets are stored in tamper-proof locations. Under this conventional point of view, cryptographic systems are modeled as black boxes, i.e., as ideal mathematical objects. Nevertheless, cryptographic systems are implemented on physical devices, which present potential side channels not considered by the security models of theoretical cryptography. In this context, fault analysis came into the scene as an alternative approach on gathering secret data.

Fault attacks are focused on attacking the physical implementation of a given cryptosystem, rather than its algorithmic structure. A fault attack relies on the principle that the cryptanalyst is allowed to manipulate the target device and induce it to abnormally operate, making it to output faulty results. These faulty results can later on be used by the cryptanalyst to derive secret information. Depending on the implementation, there are several fault induction techniques that can be deployed over the target device: exposing its surface to focused light beams, provoking variations in its power supply, exposing it to heat or radiation, inducing clock variations, among others.

The introduction of fault analysis by Boneh *et al.* [3] motivated an extensive research in the field. Since then, fault analysis has been successfully applied to disrupt standard symmetric [2] and asymmetric [1] ciphers. But there are no such results for lightweight authentication protocols, which is the intended goal of the present work.

The authors are with the Department of Electrical Engineering, University of Brasilia, Campus Darcy Ribeiro, 70910-900, Brasilia, DF, Brazil.
e-mail: {carrijo,tonicelli}@redes.unb.br, andclay@ene.unb.br
Manuscript

**Contributions.** This paper provides a fault analytic techniques against the $HB^+$ protocol. Our objective is to contribute to the specialized literature by offering an alternative perspective in the cryptanalysis of such protocols. The fault attack here proposed presents various advantages over other active attacks: (1) it does not require the eavesdropping of previous authentication procedures, (2) it can be easily adapted to disrupt other HB-variants, (3) technologically, it requires an affordable physical apparatus. Our fault analysis model assumes a great level of control over the cryptographic device. We leave as an open problem the design and application of a weaker fault analytic model.

**Organization.** This paper is organized as follows. In section II, we briefly review the protocol $HB^+$. Section III details the fault analysis model here used. Section IV describes the attacks and section V presents some performance results.

## II. DESCRIPTION OF THE PROTOCOLS

Assume that tag and reader communicate by means of a wireless insecure channel and share some piece of secret information. The $HB^+$ protocol is as follows:

**Protocol $HB^+$**

| Symbol | Meaning |
|---|---|
| $(\mathbf{x}, \mathbf{y})$ | k-bit secret key pair shared by tag and reader. |
| $v_i$ | a noise bit, such that $\Pr[v_i = 1] = \eta \in {]0, \frac{1}{2}[}$. |
| $\oplus$, and $\odot$ | bitwise XOR and inner product operations. |

1) For $i = 1$ to $r$
   a) The reader chooses a random $k$-bit string $\mathbf{a}_i \in \{0,1\}^k$ and sends it to the tag.
   b) The tag chooses a random $k$-bit string $\mathbf{b}_i \in \{0,1\}^k$ and sends it to the reader.
   c) The tag computes $z_i = \mathbf{a}_i \odot \mathbf{x} \oplus \mathbf{b}_i \odot \mathbf{y} \oplus v_i$. After computing $z_i$, the tag sends it back to the reader.
   d) The reader computes $z_i^* = \mathbf{a}_i \odot \mathbf{x} \oplus \mathbf{b}_i \odot \mathbf{y}$ and compares it to $z_i$.
2) The reader accepts the authentication as valid if $z_i^* \neq z_i$ in less than $\eta r$ rounds.

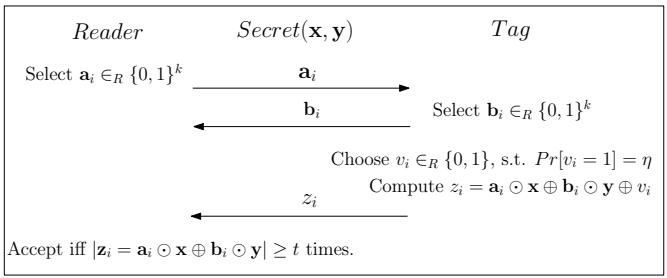

Fig. 1. Description of the protocol $HB^+$.

A legitimate tag interacting with a legitimate reader may be rejected with probability $P_{FR}$ (probability of false rejection), and an adversarial structure answering randomly at each round may be accepted with probability $P_{FA}$ (probability of false acceptance). For the $HB^+$ protocol, these probabilities are given below:

$$P_{FR} = \sum_{i=\eta r+1}^{r} \binom{r}{i} \eta^i (1-\eta)^{r-i} \text{ and } P_{FA} = \sum_{i=0}^{\eta r} \binom{r}{i} 2^{-r}$$

## III. FAULT ANALYSIS MODEL

### A. Assumed Fault Analysis Assumptions

At first, it is important to differentiate local fault injections from global fault injections. Local fault injections affect only specific regions of the device, while global fault injections affect the entire device. Thus, local fault-injection methods are more precise and can be directed to the specific device's regions that contain sensitive information. Our fault analytic model relies on local fault injections applied to the RFID tag.

In such a fault analysis model, the adversary has physical access to the device and is allowed to run it for several times while provoking faults into chosen memory areas. Specifically, we consider that the adversary is able to apply bit flipping faults to either the RAM or the internal registers of the device. Besides that, he/she can arbitrarily reset the cryptographic device and later induce other randomly chosen faults into it.

Our fault analysis assumptions are the following ones:
- The adversary is allowed to change the content of chosen memory areas to specific values, 0 or 1.
- The adversary is able to run the authentication procedure as many times he/she needs.
- The adversary knows precisely the time the faults are met.

### B. On the Possibility of Localized Fault Injections

Since we assume a strong adversarial model, it is important to conjecture about the feasibility of the underlying attack. In this field, significant results have been achieved by Hutter *et al.* [10], who offered a detailed analysis about the vulnerabilities of RFID tags to faults.

As previously stated, we assume that the adversary is able to induce faults on specific parts of the device. In [10], the authors show that it is possible to perform localized fault injections on RFID tags by means of focused laser beams. Their local fault-injection techniques relied on an affordable physical apparatus: an optical microscope equipped with an integrated incident illumination device, and a laser diode mounted on the top of the microscope camera port.

Remarkably, the authors report that the microscope provided the capability of exploring the device's internal structure very accurately. They describe that their method allowed them to interfere data, control lines, memory blocks and driver circuits.

Thus, it is possible to conjecture that our fault analytic model, despite being strong, is not far away from reality.

## IV. DESCRIPTION OF THE ATTACK

Our method consists of performing fault insertions into the device and trying to authenticate it on a legitimate reader. Based on the result of the authentication procedure, the cryptanalyst infers the actual value of the modified bit. The inherent beauty of this attack resides in the fact that the adversary is not required to eavesdrop any previous authentication procedures.

Let **w** denote the memory area that stores the binary string composed by the concatenation of the two secret keys, i.e., $\mathbf{w} = \mathbf{x}||\mathbf{y}$, such that $\mathbf{w} \in \{0,1\}^{2k}$.

The procedure $\text{FLIP}(\mathbf{w}[i], b)$ changes the memory content of $\mathbf{w}[i]$ to the binary value $b$.

---

**Algorithm 1** $\text{FLIP}_i(\mathbf{w}, b)$: Flip the $i$-th bit of the string **w** to $b$, such that $b \in \{0,1\}$.

---

**Require:** **w** such that $\mathbf{w} \in \{0,1\}^m$.
**Ensure:** $\overline{\mathbf{w}}$, such that $\overline{\mathbf{w}}[i] = b$, and for all $j \neq i$, $\overline{\mathbf{w}}[j] = \mathbf{w}[j]$.

---

The function $\text{HBplusAuthentication}(\mathbf{v})$ denotes the process of trying to authenticate by using a $2k$-binary string **v**. The function $\text{HBplusAuthentication}(\mathbf{v})$ returns TRUE if the authentication procedure succeeds, or returns FALSE if the authentication procedure fails.

Algorithm 2 describes the attack on $\text{HB}^+$.

---

**Algorithm 2** BreakHBplus(**w**,q): Compute the two shared secret keys **x** and **y**.

---

**Require:** Memory area **w** and parameter $q \in \mathbb{N}^*$.
**Ensure:** The secret key pair $(\mathbf{x}, \mathbf{y})$, such that $(\mathbf{x}, \mathbf{y}) \in \{0,1\}^k \times \{0,1\}^k$. The secret key pair is correctly discovered with high probability.

**for** $i = 1$ to $2k$ **do**
  $\text{FLIP}(\mathbf{w}[i], 0)$
  $cont \leftarrow 0$
  **for** $j = 1$ to $q$ **do**
    **if** $\text{HBplusAuthentication}(\mathbf{w})$ **then**
      $counter \leftarrow counter + 1$
    **end if**
  **end for**

  **if** $counter \geq q/2$ **then**
    $\text{ExtractedKey}[i] \leftarrow 0$
  **else**
    $\text{ExtractedKey}[i] \leftarrow 1$
    $\text{FLIP}(\mathbf{w}[i], 1)$
  **end if**
**end for**

$\mathbf{x} \leftarrow \text{ExtractedKey}[1, \ldots, k]$
$\mathbf{y} \leftarrow \text{ExtractedKey}[k+1, \ldots, 2k]$

**return** $(\mathbf{x}, \mathbf{y})$

---

## V. PERFORMANCE RESULTS

### A. Information Theoretic Measures

Prior to evaluating the effectiveness of the attack against the protocol $\text{HB}^+$, we shall define some information theoretic quantities that will be useful in our analysis. Particularly, we are interested in two tasks: (1) quantifying the correlation between the key stored in the device and the key extracted by the attacker and (2) characterizing the level of information leaked by executing the underlying cryptanalytic method.

Let $X$ and $Y$ be discrete random variables defined over finite alphabets $\mathcal{X}$ and $\mathcal{Y}$, respectively. The definitions are as follow:

The entropy of $X$ can be visualized as the amount of uncertainty contained in it and is given by:

$$H(X) = -\sum_{x} P_X(x) \log_2 P_X(x), \quad \text{where } 0 \leq H(X) \leq \log_2 |\mathcal{X}|.$$

Loosely speaking, the equivocation of $X$ given $Y$ measures the amount of remaining uncertainty on $X$ given that $Y$ is known. Its definition is given next:

$$H(X|Y) = -\sum_{x,y} P_{XY}(x,y) \log_2 P_{X|Y}(x,y),$$

Where $0 \leq H(X|Y) \leq H(X)$.

The mutual information between $X$ and $Y$ measures the correlation between them and is defined as:

$$I(X;Y) = \sum_{x,y} P_{XY}(x,y) \log_2 \frac{P_{XY}(x,y)}{P_X(x) P_Y(y)}$$

Where $0 \leq I(X;Y) \leq H(X)$.

Additionally,

$$I(X;Y) = H(X) - H(X|Y).$$

### B. Results

**Probability of Failure.**

In the first place, we shall calculate the probability of committing an error when retrieving a single bit of a shared secret key. This probability of error can be trivially derived.

*Lemma 1:* Let $\mathbf{w} \in \{0,1\}^{2k}$ denote a uniformly distributed secret key stored in a RFID tag, and let $\mathbf{w}'_q \in \{0,1\}^{2k}$ denote the key guessed by an attacker who executes $q$ queries to the device. The probability of error, i.e., the probability of retrieving the bit $\mathbf{w}'_q[i] = \overline{b}$ when the actual bit is $\mathbf{w}[i] = b$, is given by:

$$P_e(q) = \sum_{i=\lfloor q/2 \rfloor + 1}^{q} \binom{q}{i} (p)^i (1-p)^{q-i}$$

Where $p = \Pr(\mathbf{w}'_1[i] \neq \mathbf{w}[i]) = \frac{1}{2}(P_{FA} + P_{FR})$.

*Proof:* Initially, we obtain the probability of error when one single query ($q = 1$) is executed by the attacker.

$$p = \sum_{b \in \{0,1\}} \Pr(\mathbf{w}[i] = b) \cdot \Pr(\mathbf{w}'_1[i] = \overline{b} \,|\, \mathbf{w}[i] = b)$$
$$= \frac{1}{2} \cdot \Pr(\mathbf{w}'_1[i] = 1 \,|\, \mathbf{w}[i] = 0) + \frac{1}{2} \cdot \Pr(\mathbf{w}'_1[i] = 0 \,|\, \mathbf{w}[i] = 1)$$

- $\Pr(\mathbf{w}'_1[i] = 1 \,|\, \mathbf{w}[i] = 0)$.
  In this case, the adversary injects a zero into the position $\mathbf{w}[i]$. Consequently, the attacker's action did not alter the value of the stored secret key and the tag remains valid. An error occurs if the reader rejects the tag.

$$\Pr(\mathbf{w}'_1[i] = 1 \,|\, \mathbf{w}[i] = 0) = P_{FR}$$

- $\Pr(\mathbf{w}'_1[i] = 0 \,|\, \mathbf{w}[i] = 1)$.
  After the fault injection, the value of the stored secret key is modified, and the tag is no longer valid. Consequently, the reader is expected to reject the tag. An error occurs if the reader authenticates the tag.

$$\Pr(\mathbf{w}'_1[i] = 1 \,|\, \mathbf{w}[i] = 0) = P_{FA}$$

Therefore, $p = \frac{1}{2}(P_{FA} + P_{FR})$.

In the next step we calculate the probability of error when an arbitrary number $q$ of queries is executed, which is given by:

$$P_e(q) = \sum_{i=\lfloor q/2 \rfloor + 1}^{q} \binom{q}{i} (p)^i (1-p)^{q-i}$$

Figure 2 shows the graphic of $p = f(\eta, r)$. We may conclude that the more reliable is the authentication procedure, the more reliable is the fault analytic method. Furthermore, it is easily observable that $P_e(q)$ asymptotically goes to zero for sufficiently large $q$. ∎

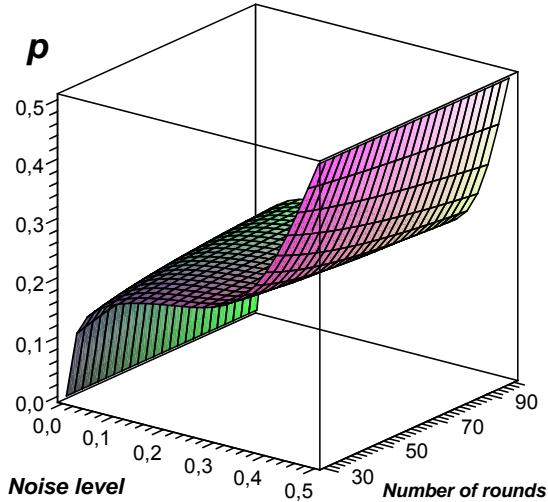

Fig. 2. Probability of error as a function of $\eta$ (the noise level) and $r$ (number of rounds).

**Information Leakage.**

We analyze the effectiveness of the attack by using the definitions in section V-A. We also empirically demonstrate that $P_e(q)$ approaches zero for a sufficiently large number of queries.

One can see that the process of extracting one isolated bit of the secret key is analogous to the process of transmitting a single bit over a binary symmetric channel (BSC) with crossover probability $p(\eta, r)$. Let $h(p) = -p \log p - (1-p) \log(1-p)$ denote the binary entropy function, for the specific case where the key is uniformly distributed, we have that $H(\mathbf{w}[i] \,|\, \mathbf{w}'_q[i]) = h(P_e(q))$ and $I(\mathbf{w}[i]; \mathbf{w}'_q[i]) = 1 - h(P_e(q))$.

We can observe that $H(\mathbf{w}[i] \,|\, \mathbf{w}'_q[i]) \to 0$ and $I(\mathbf{w}[i]; \mathbf{w}'_q[i]) \to 1$ for sufficiently large $q$. This is illustrated in the next tables.

Furthermore, we should point out that the underlying attack requires the injection of $2k$ faults and the realization of $2kq$ authentication procedures. Thus it presents a linear time complexity $O(k)$.

| $q$ | $P_e(q)$ | $H(\mathbf{w}[i] \,|\, \mathbf{w}'_q[i])$ | $I(\mathbf{w}[i]; \mathbf{w}'_q[i])$ |
|---|---|---|---|
| 7 | 0.0289 | 0.2943 | 0.8111 |
| 11 | 0.0094 | 0.1888 | 0.9227 |
| 17 | 0.0019 | 0.772 | 0.9800 |
| 19 | 0.0011 | 0.0199 | 0.9873 |

TABLE I
RESULTS FOR PARAMETERS $\eta = 0.125$, $r = 40$, $p(0.125, 40) = 0.1919$.

| $q$ | $P_e(q)$ | $H(\mathbf{w}[i] \,|\, \mathbf{w}'_q[i])$ | $I(\mathbf{w}[i]; \mathbf{w}'_q[i])$ |
|---|---|---|---|
| 7 | 0.0384 | 0.2348 | 0.7651 |
| 11 | 0.0143 | 0.1080 | 0.8919 |
| 17 | 0.0035 | 0.0334 | 0.9666 |
| 19 | 0.0022 | 0.0225 | 0.9775 |

TABLE II
RESULTS FOR PARAMETERS $\eta = 0.125$, $r = 80$, $p(0.125, 80) = 0.2084$.

| $q$ | $P_e(q)$ | $H(\mathbf{w}[i] \,|\, \mathbf{w}'_q[i])$ | $I(\mathbf{w}[i]; \mathbf{w}'_q[i])$ |
|---|---|---|---|
| 7 | 0.0462 | 0.2702 | 0.7298 |
| 11 | 0.0187 | 0.1341 | 0.8656 |
| 17 | 0.0051 | 0.0465 | 0.9535 |
| 19 | 0.0033 | 0.0326 | 0.9674 |

TABLE III
RESULTS FOR PARAMETERS $\eta = 0.25$, $r = 80$, $p(0.25, 80) = 0.2201$.

[6] H. Gilbert, M. Robshaw, H. Seurin. $HB^{\#}$: Improving the Security and Efficiency of $HB^{+}$. Advances in Cryptology – EUROCRYPT 2008, Lecture Notes in Computer Science vol. 4965, Springer-Verlag, pages 361–378, 2008.

[7] H. Gilbert, M. Robshaw, H. Seurin. Good Variants of $HB^{+}$ are Hard to Find. 2008 Financial Cryptography Conference, Lecture Notes in Computer Science vol. 5143, Springer-Verlag, pages 156–170, 2008.

[8] H. Gilbert, M. Robshaw, H. Silvert. An active attack against $HB^{+}$ - a provable secure lightweight protocol. Cryptology ePrint Archive, Report 2005/237, 2005, available at http://eprint.iacr.org/2005/237.pdf.

[9] N. J. Hopper and M. Blum. Secure Human Identification Protocols. Advances in Cryptology – ASIACRYPT 2001, Lecture Notes in Computer Science vol. 2248, Springer-Verlag, pages 52–66, 2001.

[10] M. Hutter, J.-M. Schimidt and T. Plos. RFID and Its Vulnerability to Faults. CHES 2008, Lecture Notes in Computer Science vol. 5154, Springer-Verlag, pages 363–379, 2008.

[11] A. Juels and S. A. Weis. Authenticating pervasive devices with Human Protocols. Advances in Cryptology – CRYPTO 2005, Lecture Notes in Computer Science vol. 3621, Springer-Verlag, pages 293–308, 2005.

[12] J. Munilla and A. Peinado. A further step in the HB-family of lightweight authentication protocols. Computer Networks, vol. 51, Elsevier, pages 2262–2267, 2007.

[13] H. Ouafi, R. Overbeck and S. Vaudenay. On the Security of $HB^{\#}$ against a Man-in-the-Middle Attack. Advances in Cryptology – ASIACRYPT 2008, Lecture Notes in Computer Science vol. 4965, Springer-Verlag, pages 361–378, 2008.

[14] Y. Seurin. Primitives et protocoles cryptographiques à sécurité prouvée. PhD thesis. Université de Versailles Saint-Quentin-en-Yveline, 2009. Available at http://yannickseurin.free.fr/pubs/these_Yannick_Seurin.pdf